%% file: main.tex
\pdfoutput=1
\documentclass[amsmath,amssymb,aps,superscriptaddress,twocolumn,floatfix]{revtex4-1}
\usepackage{graphicx} 
\usepackage{wrapfig}
\usepackage{bm}
\usepackage[usenames]{color}
\usepackage{MnSymbol}
\usepackage{hyperref} 
\hypersetup{
    colorlinks=true,
    linkcolor=blue,
    filecolor=magenta,      
    urlcolor=cyan,
    pdftitle={Overleaf Example},
    pdfpagemode=FullScreen,
    }
\usepackage{subfigure}
\usepackage[normalem]{ulem}

\newcommand{\head}[1]{#1}
\newcommand{\tail}[1]{#1}
\definecolor{darkgreen}{RGB}{29, 117, 2}
\definecolor{darkblue}{RGB}{44, 92, 176} 
\definecolor{darkred}{RGB}{199, 46, 36} 
\definecolor{grey}{rgb}{.6,.6,.6}
\newcommand{\revision}[2]{#2}

\newcommand{\sq}{\langle S_q(\vec{k})\rangle}
\begin{document}

\title{Quantum Phases of Transition Metal Dichalcogenide Moir{\'e} Systems}
\author{Yiqing Zhou}
\affiliation{Laboratory of Atomic and Solid State Physics, Cornell University, Ithaca, NY 14853, USA}
\author{D. N. Sheng}
\affiliation{Department of Physics and Astronomy, California State University, Northridge, CA 91330, USA}
\author{Eun-Ah Kim}
\affiliation{Laboratory of Atomic and Solid State Physics, Cornell University, Ithaca, NY 14853, USA}

\begin{abstract}
Moir\'e systems provide a rich platform for studies of strong correlation physics. Recent experiments on hetero-bilayer transition metal dichalcogenide (TMD) Moir\'e systems are exciting in that they manifest a relatively simple model system of an extended Hubbard model on a triangular lattice.  
Inspired by the prospect of the hetero-TMD Moir\'e system's potential as a solid-state-based quantum simulator, we explore the extended Hubbard model on the triangular lattice using the density matrix renormalization group (DMRG). 
Specifically, we explore the two-dimensional phase space \revision{of the kinetic energy relative to the on-site Coulomb interaction strength $t/U$ and the further-range interaction strength $V_1/U$}{spanned by the key tuning parameters in the extended Hubbard model, namely, the kinetic energy strength and the further-range  Coulomb interaction strengths}. We find competition between Fermi fluid, chiral spin liquid, spin density wave, and charge order. In particular, 
our finding of the optimal further-range interaction for the chiral correlation presents a tantalizing possibility.
\end{abstract}
\maketitle
The triangular lattice Hubbard model has long been of intense interest \cite{Sahebsara2008Phys.Rev.Lett., Yoshioka2009Phys.Rev.Lett., Yang2010Phys.Rev.Lett., Mishmash2015Phys.Rev.B, Shirakawa2017Phys.Rev.B, Venderley2019Phys.Rev.B., Szasz2020Phys.Rev.X, Zhu2020, 
Song2020, Gannot2020, Szasz2021, Chen2021, Wietek2021, Peng2021}, since the geometric frustration and quantum fluctuation can lead to a rich set of possibilities.
In particular, previous density matrix renormalization group (DMRG) studies have shown a robust metal-insulator transition (MIT)  at half-filling~\cite{Mishmash2015Phys.Rev.B}. More recently, the possibility of the chiral spin liquid (CSL) phase preempting the MIT~\cite{Szasz2020Phys.Rev.X, Chen2021} has been predicted. Other possible ordered phases have been proposed~\cite{Wietek2021}.  \revision{However,  the lack of continuous experimental control over the ratio of the interaction strength and the bandwidth, $U/t$, the key tuning parameter in theoretical studies,  has made it challenging to test these predictions. }{However, the well-studied standard Hubbard model dismisses the long-range interactions which are ubiquitous in materials explored in experiments; therefore, predictions made by previous works are challenging to test in realistic experimental setups. }

Following the proposal of \cite{Wu2018Phys.Rev.Lett.} that hetero-bilayer transition metal dichalcogenide (TMD) Moir{\'e} systems can realize the triangular lattice Hubbard model~\cite{Kennes2021Nat.Phys.}, recent experiments on hetero-TMD Moir{\'e} systems have indeed observed Mott insulating states at half-filling \cite{Regan2020Nature, Tang2020Nature}. Furthermore, continuous control over \revision{$U/t$}{hopping strength relative to the interaction strength} is now accessible~\cite{Li2021, Ghiotto2021}. However, as evidenced by the charge order at fractional fillings~\cite{Xu2020Nature, Jin2020} the TMD systems have further-range interactions due to a low charge density and resulting low screening. Excitingly, these further-range interactions can also be tuned in experiments~\cite{Li2021}, presenting a two-dimensional space, spanned by \revision{$U/t$ and $V_1/U$}{the hopping and the further range interaction strengths with respect to the on-site Coulomb interaction strength}, to explore a plethora of quantum phases. In comparison, computational investigations have so far been restricted to systems with only on-site interactions\revision{}{, confining us to a one-dimensional phase space controlled purely by ratio of on-site Coulomb interaction and hopping strength}. 

\begin{figure}
  \centering
  \includegraphics[width=0.85\columnwidth]{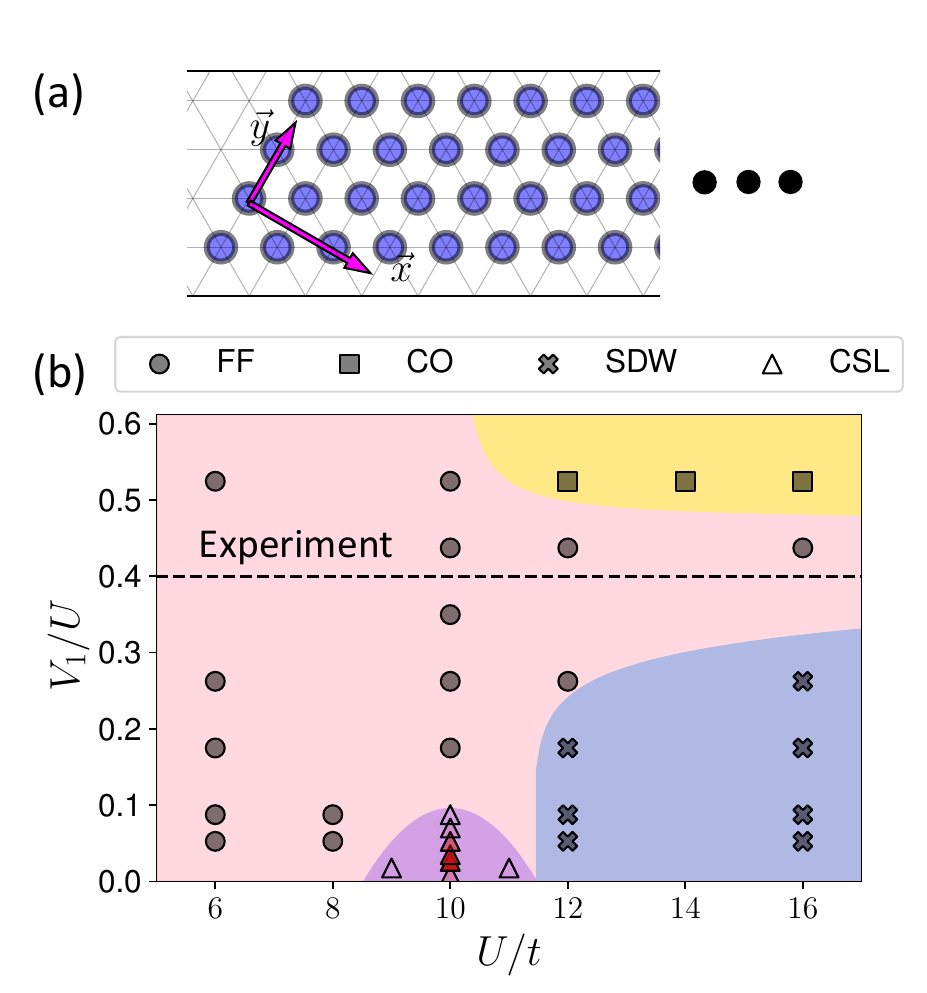}
  \vspace{-.5cm}
  \caption{
  The system under study and the phase diagram. (a) A partial sketch of a YC4 cylinder used in the DMRG calculation, where $x,y$ directions are specified by the arrows. The number of lattice sites along $y$ direction $L_y=4$. The boundary condition is periodic in $y$ and open along the cylinder axis direction. 
  (b)The phase diagram of the extended Hubbard model on a triangular lattice. 
  The symbols represent phases observed using DMRG; the color shadings are the schematic extents of the phases.
  The color intensity of the CSL markers (triangles) represents the strength of chiral ordering.\revision{}{The dashed line marks the $V_1/U$ accessed by an instance of the experimental setup.}}
  \label{fig:sketch_phase_diagram}
\end{figure}

Motivated by the experimental developments, in this work, we explore the extended Hubbard model on a triangular lattice and study the phase space upon tuning the ratio of hopping to on-site interaction $t/U$ and the relative strengths  of long-range interactions \revision{}{parameterized by} $V_1/U$  (Fig.\ref{fig:sketch_phase_diagram}(b)). Using DMRG, we make the first pass over this large phase space and benchmark our results against the standard Hubbard model limit \cite{Mishmash2015Phys.Rev.B, Shirakawa2017Phys.Rev.B, Szasz2020Phys.Rev.X, Chen2021}. By studying the effect of further-range interactions, we study competition between charge order, chiral spin liquid, and spin density wave at half-filling. The rest of the paper is organized as follows: We first introduce the extended Hubbard model under study and identify a rich phase diagram with Fermi fluid (FF), chiral spin liquid (CSL), and spin density wave (SDW) phases in weak long-range interaction region,  showing consistency between the extended Hubbard model in small $V_1/U$ limit and the standard Hubbard model. 
Interestingly, we find that intermediate long-range interaction can enhance the chiral order. 
Finally, we discuss the charge-ordered states (CO) promoted by stronger long-range interactions. We close with a summary of the quantum phases we observe and provide future outlooks.

{\it Model--} We consider the following extended Hubbard model on a triangular lattice,
\begin{equation}
\label{Eq:Hamiltonian}
  \begin{split}
    H = & -t\sum_{\left<ij\right>}\left(c_i^\dag c_j + H.c.\right)  
    + U\sum_{i} n_{i, \uparrow} n_{i, \downarrow} \\
    & + V_1 \sum_{\left<ij\right>} n_i n_j + V_2 \sum_{\langle\!\langle ij \rangle\!\rangle } n_i n_j + V_3 \sum_{\langle\!\langle\!\langle ij\rangle\!\rangle\!\rangle} n_i n_j,
  \end{split}
\end{equation}
where $t, U, V_i$ represnt hopping, on-site interaction, $i$-th nearest neighbor interaction strength respectively. 
\revision{We choose $V_2/V_1\approx 0.357$ and $V_3/V_1 \approx 0.260$ to follow an hetero-bilayer TMD experiment setup \cite{Li2021}  (derivation shown in Supplemental Materials~\cite{sm}). }{We follow an hetero-bilayer TMD experiment setup \cite{Li2021} and set $V_2/V_1\approx 0.357$ and $V_3/V_1 \approx 0.260$   (derivation shown in Supplemental Materials~\cite{sm}). 
The choice of $V_2/V_1$ and $V_3/V_1$ does not quantitatively change the phases observed in the phase space within an experimentally feasible parameter range (see Supplemental Materials~\cite{sm} for results comparing different $V_2/V_1$ and $V_3/V_1$ ratios).}
Therefore, the relative strength of further-range interactions is tuned by the single parameter $V_1/U$.
Given the independent experimental control over the bandwidth and the range of interaction, we explore the ground state in a phase space spanned by the on-site interaction strength, $U/t$ and the further-range interaction strength represented by $V_1/U$. With the MIT in mind, we focus on the 
half-filling state at zero total spin $S=0$. 
We perform large-scale DMRG calculations on YC4 cylinders as sketched in  Fig.~\ref{fig:sketch_phase_diagram}(a). 
\revision{}{Note that the YC$n$ (XC$n$) cylinders have one of the lattice edge parallel (perpendicular) to the periodic direction and have $n$ sites ($n/2$ unit cells) along the periodic direction. To investigate the effect of further-range interaction on the CSL phase, we focus on YC geometry in which previous studies on standard triangular Hubbard model have observed that chiral phase is easier to observe \cite{Szasz2020Phys.Rev.X, Chen2021} as opposed to the XC geometry. }
The YC4 geometry can lead to double counting in the second and third nearest neighboring sites; we avoid the double counting by coupling each pair of sites only once.
We compare various cylinder lengths, $L_x=16, 32, 48$, and keep $10,000$ to $30,000$ \cite{mcculloch2002EPL} states  to get high-accuracy numerical results \revision{}{(See Supplemental Materials~\cite{sm} for more details)}. 

We observe a rich phase space in a parameter space reachable by existing experimental devices. 
As shown in Fig. \ref{fig:sketch_phase_diagram}(b), the phases we observe include chiral spin liquids (CSL), spin density waves (SDW), charge-ordered states (CO), and Fermi fluids (FF), with
the long-range interactions further enriching the phase space compared to the standard Hubbard model. 
In the absence of further-range interactions, the system in the large $U/t$ limit can be well-captured by an effective spin model~\cite{MacDonald1988Phys.Rev.B.} with antiferromagnetic nearest neighbor interaction driven by super exchange on the triangular lattice. The quantum fluctuation in intermediate $U/t$ with the geometric frustration of the triangular lattice might drive CSL\cite{Cookmeyer2021}.
The further-range interaction $V_1$ can impact the ground state both in the spin  and charge sectors.
 First,  in the small $V_1/U$ limit,  $V_1$ interaction  enhances the spin exchange interactions by a factor $U/(U-V_1)$, while it can also be mapped to ferromagnetic direct exchange interaction between nearest neighbors, suppressing antiferromagnetic ordering tendencies. 
This expectation is borne out in our discovery that the chiral order peaks at a finite small $V_1/U$ in the intermediate $U/t$ range as highlighted by the color intensity in Fig.~\ref{fig:sketch_phase_diagram}(b).
Second, as the long-range interaction strength increases, charge order is promoted, and spin ordering is suppressed. 
Thus we see the melting of CSL and SDW and then the emergence of CO as $V_1/U$ increases\revision{}{\cite{caprara1997spin}}.
\tail{The richness of the phase space demonstrates the importance of long-range interactions.}
Beyond the intuitive understanding of these phases, we present detailed numerical evidence for each phase below.

\begin{figure}[b!]
  \centering
  \includegraphics[width=0.85\columnwidth]{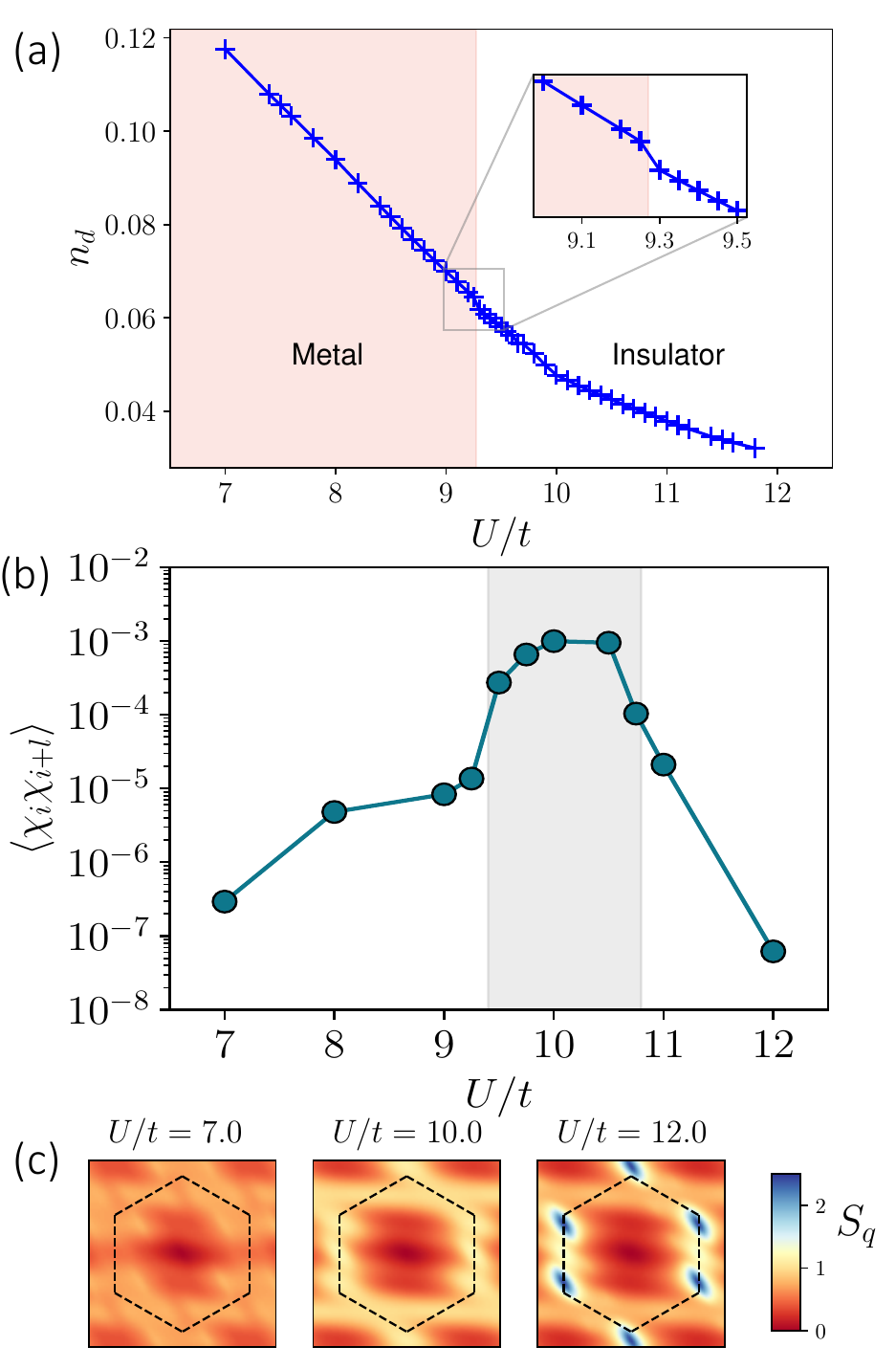}
  \vspace{-0.5cm}
  \caption{
    Small $V_1/U$ limit.
       (a) Double occupancy $n_d$ as a function of $U/t$ at $V_1/U \approx 0.0175$. 
    The discontinuous drop highlighted in the inset shows the transition from a metallic to an insulating phase.
    Data is from $4 \times 16$ YC cylinder calculations.
    (b) Chiral correlation function $\langle \chi_i\chi_{i+l} \rangle$ versus $U/t$ at $V_1/U \approx 0.0175$ in a log scale. As a guide to the eye, the grey shaded region marks the range of $U/t$ where CSL lives. 
    (c) Spin structure factors $\sq$ plotted in momentum space at $V_1/U \approx 0.0175$. 
    Three panels correspond to $U/t =7.0, 10.0$ and $12.0$ from left to right, respectively. 
    The black dashed lines mark the boundary of the Brillouin zone. Data is from $4 \times 32$ YC cylinders for (b-c). 
  }
  \label{fig:small_v}
\end{figure}

{\it Small $V_1/U$ limit--} \head{In the small $V_1/U$ limit ($V_1/U\approx0.0175$), we reproduce the latest results of the Hubbard model with $V_1/U=0$~\cite{Shirakawa2017Phys.Rev.B,Szasz2020Phys.Rev.X,Chen2021}}.
First, we detect the MIT by calculating the double occupancy,
\(n_d = \langle n_{\uparrow}n_{\downarrow}\rangle\). 
It shows a  discontinuous drop at the critical value of $(U/t)_\text{MIT}\sim9.2$, 
upon increasing $U/t$ (Fig.~\ref{fig:small_v}(a)). 
 This observation is consistent with an earlier study~\cite{Chen2021}, which reports the MIT at $U/t \approx 9.0$ for the Hubbard model.

\begin{figure}[h]
  \centering
  \includegraphics[width=0.86\columnwidth]{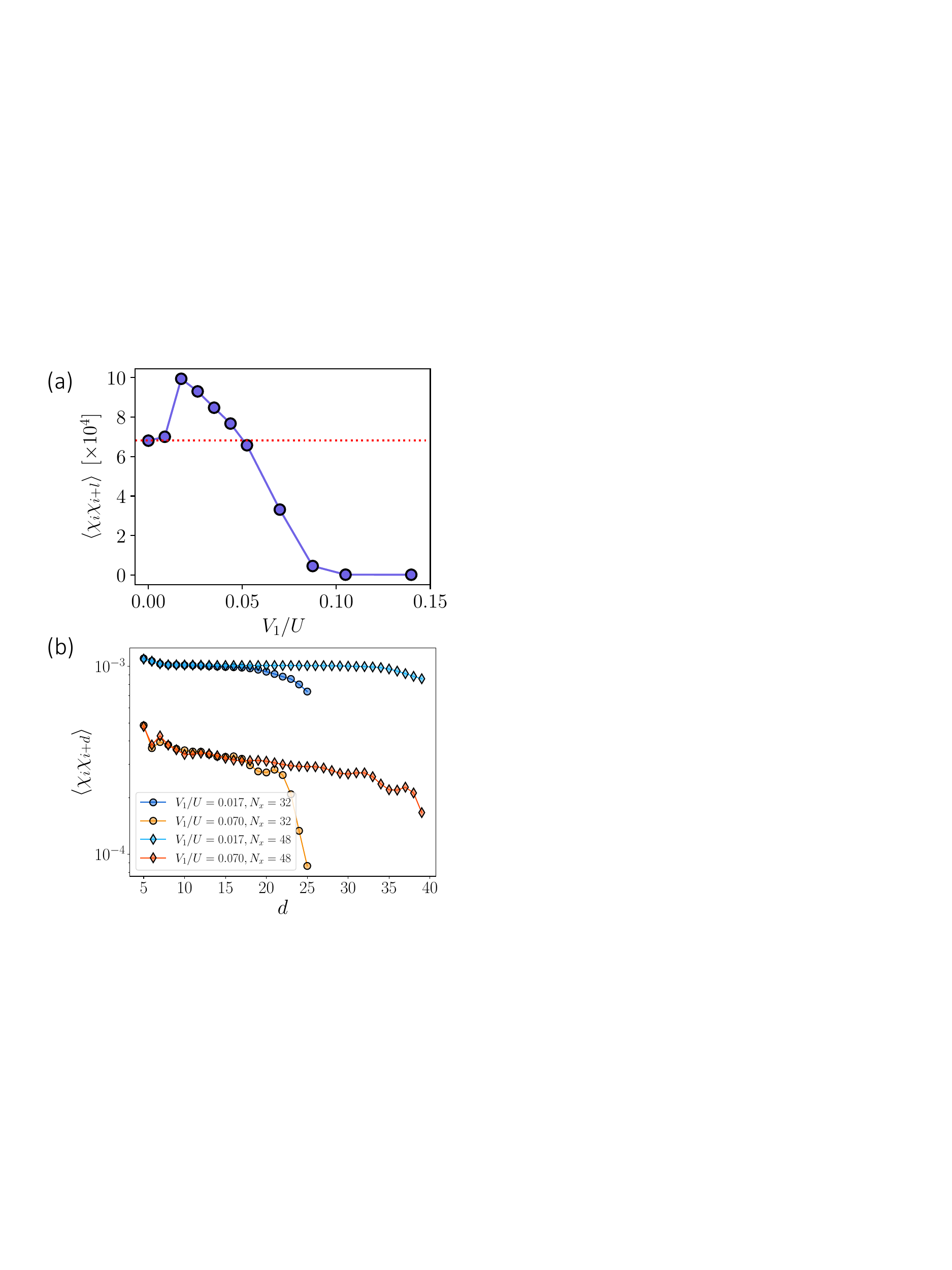}
  \vspace{-0.5cm}
  \caption{
        Chiral correlation function.
       (a) Chiral correlation function $\langle \chi_i\chi_{i+l} \rangle$ versus $V_1/U$ at $U/t=10.0$.
        The red dashed horizontal line denotes $\langle \chi_i\chi_{i+l} \rangle$ at $V_1/U=0$.
         Data points are obtained from $4 \times 32$ cylinders, and the site separation is chosen to be by half the cylinder length, $l=L_x/2=16$. 
        (b) Chiral correlation $\langle \chi_i\chi_{i+d} \rangle$ as a function of site separation $d$ with fixed $U/t=10.0$ and different $V_1/U$. Results from two systems sizes, $L_y=4, L_x=32$ and $48$ are presented.
  }
  \label{fig:csl}
\end{figure}

Next, we identify the above metal-insulator transition to take the system into a CSL phase by calculating the spin structure factor (Fig.~\ref{fig:small_v}(c)) and the chiral correlation function (Fig.~\ref{fig:csl}).
We consider the chiral correlation function  $\langle \chi_m \chi_n \rangle$, where the chiral order parameter $\chi_i = \vec{S}_i \cdot \left( \vec{S}_j \times \vec{S}_k \right)$ is defined on a triangle centered around lattice site $i$, involving all three corner sites $i,j,k$.
\footnote{We consider the chiral correlation function instead of the chiral order parameter because we use real wavefunction for the ground state where the $\langle \chi_i \rangle=0$ due to time-reversal symmetry.}
We observe, in Fig.~\ref{fig:small_v}(b)  between $U/t \approx 9.4$ and $10.8$, a long-range chiral correlation $\langle \chi_i \chi_j\rangle$ that is orders of magnitude higher than that in neighboring phases. 
The lower boundary of the long-range chiral correlation $U/t \approx 9.4$ matches where the MIT is observed from the double-occupancy (see Fig.~\ref{fig:small_v}(a).
In addition, 
the featureless spin structure factor, \(\langle S_q(\vec{k})\rangle = \frac{1}{N}\sum_{ij}  e^{i \vec{k} \cdot \vec{r}_{ij}} \langle \vec{S_i} \cdot \vec{S_j} \rangle \) confirms absence of magnetic order associated with the intermediate $U/t$ insulating phase.
Note that the slight anisotropy is caused by the finite-size geometry of the YC4 cylinders. We also analyze real space correlation function in Supplemental Materials~\cite{sm}.
 Observing  long-range chiral correlations that are at the same order of magnitude as reported in Refs.~\cite{Szasz2020Phys.Rev.X, Chen2021} for the Hubbard model with $V=0$ \footnote{The chiral order parameter $\chi$ in this work is defined in terms of the spin operators, which differs from the definition using Pauli matrices \cite{Szasz2020Phys.Rev.X} by a factor of $1/8$.}, we conclude that the insulating state immediately after the MIT is CSL. 

Increasing $U/t$ further,  we observe an additional phase transition within the insulating phase. Upon this second transition, the spin structure factor gains sharp peaks at corners of the Brillouin zone (see Fig.~\ref{fig:small_v}(c)). Similar peaks in $\sq$ have been observed in the Hubbard model~\cite{Shirakawa2017Phys.Rev.B, Chen2021}. 
These peaks distinguish the large  $U/t$ insulating phase from the intermediate $U/t$ CSL phase and evidence the emergence of spin density waves with wave vectors $\mathbf{Q}_{\text{SDW}}= (\sqrt{3}\pi/2a, \pi/2a)$ and $(\sqrt{3}\pi/2a, -\pi/2a)$. At the same time, the chiral correlation dives down, indicating the non-chiral nature of the spin density wave state (see Fig.~\ref{fig:small_v}(b)). 


{\it Effects of further-range interactions --} We study phase transitions driven by the further-range interactions, fixing the on-site interaction strength at $U/t=10$, which is close to the center of the CSL phase. The further-range interaction has a non-monotonic effect on the CSL phase as it is shown through the chiral correlation at half the length of the YC4 cylinder, $l=L_x/2$  as a function of $V_1/U$ in Fig.~\ref{fig:csl}(a).
The non-monotonicity comes with the dramatic enhancement in the chiral correlation at intermediate $V_1/U$ compared to the original Hubbard model. Upon further increasing  $V_1/U$, the system leaves the sweet spot, and chiral correlation dies out at $V_1/U \gtrsim 0.07$.  
Within the CSL phase, we investigate the chiral correlation strength as a function of distance $d$.
As shown in Fig.~\ref{fig:csl}(b), we find the chiral correlation to be long-ranged, nearly constant as a function of $d$ in the CSL phase. We present results from systems with $L_x=32$ and $48$. Despite the finite-size effects near the end of the cylinder, both systems have consistent chiral correlation strength. The region of nearly constant correlations grows with $L_x$, indicating true chiral order at large $L_x$ limit.

\begin{figure}[h]
  \centering
  \includegraphics[width=0.98\columnwidth]{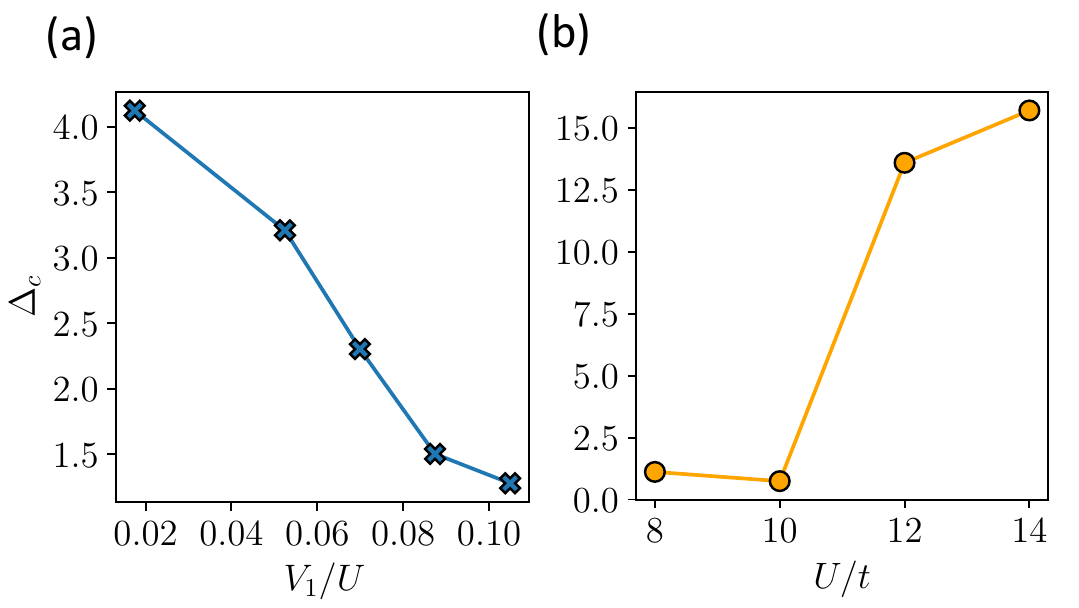}
  \caption{
   Charge gap. (a) Charge gap $\Delta_c$ along a vertical cut in phase diagram at $U=10$ with varying further-range interaction strength parameterized by $V_1/U$. 
   (b) Charge gap along a horizontal cut in phase diagram at $V_1/U \approx 0.6119$ with varying $U/t$. 
  }
  \label{fig:csl-charge-gap}
\end{figure}

\begin{figure}[htbp]
  \centering
  \includegraphics[width=0.95\columnwidth]{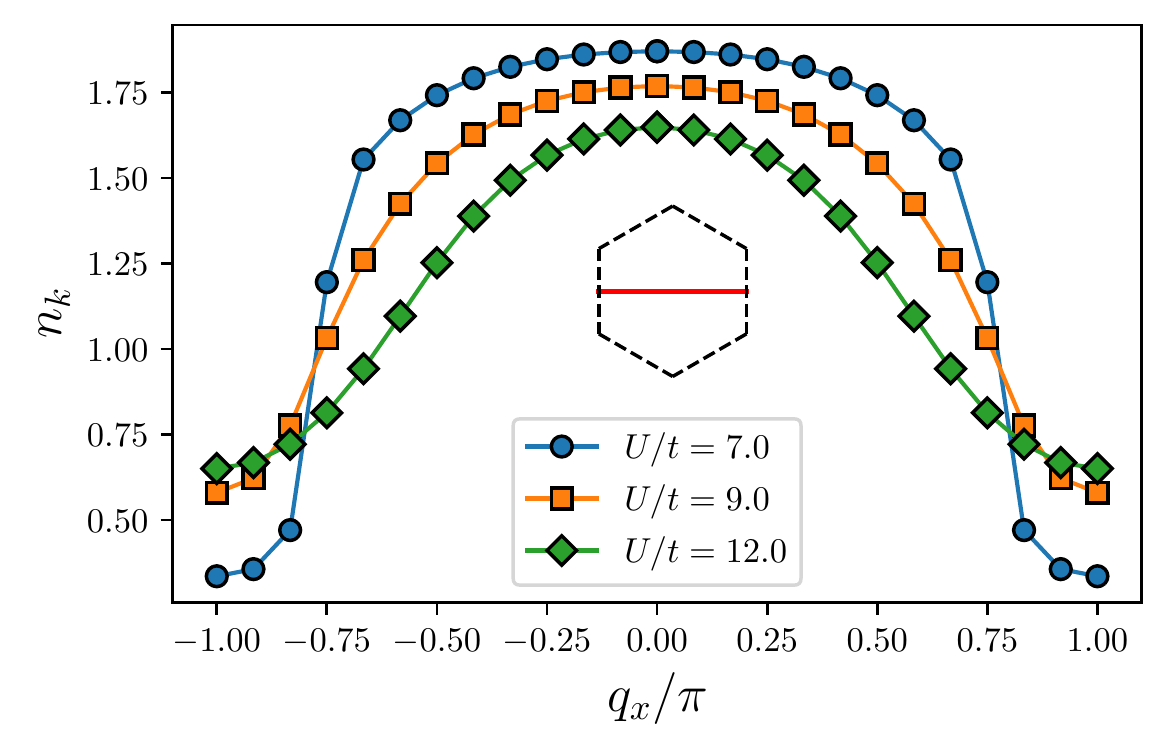}
  \caption{
   Electron momentum distribution $n_k$ along a momentum space cut at fixed further-range interaction strength $V_1/U=0.0175$ and various on-site interaction strengths $U/t= 7.0 ,9.0$ and $12.0$. The inset shows a Brillouin zone with the cut marked by the red line.
  }
  \label{fig:ff}
\end{figure}

\begin{figure*}[]
  \centering
  \includegraphics[width=1.9\columnwidth]{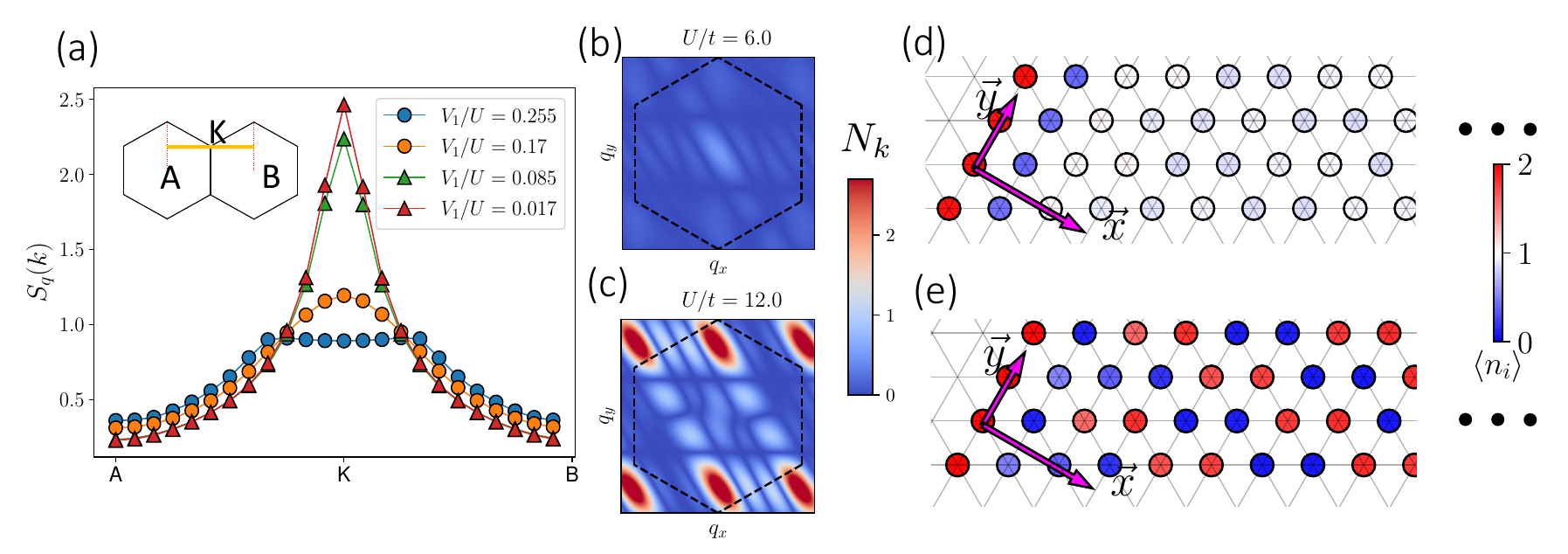}
  \caption{
    Large $U/t$ limit. 
    (a) Spin structure factor $S_q(k)$ with  $U/t=12.0$ at various $V_1/U$ along a cut in momentum space highlighted by the yellow line in the inset. The peaks are observed around the $K$ point.
    Panels (b)-(e) show charge density at $V_1/U\approx 0.612$. 
    (b) Charge density in momentum space $N_k$ for $U/t=6.0$.
    (c) $N_k$ for $U/t=12.0$. 
    The black dashed lines represent the boundary of the Brillouin zone. 
    (d) Charge density in real space $n_i$ for $U/t=6.0$.
    (e) $n_i$ for $U/t=12.0$. 
  }
  \label{fig:large_u}
\end{figure*}



As $V_1/U$ increases, the CSL melts into a Fermi fluid phase. 
\revision{}{As shown in Fig.~\ref{fig:csl-charge-gap}, we look at the charge gap $\Delta_c$ along a vertical line in the phase diagram (Fig~\ref{fig:sketch_phase_diagram}) at $U/t=10$. The charge gap diminishes as the further-range interaction strength increases, evidencing the melting of CSL insulating phase to a  FF phase.}
This FF phase fills in the large space between CSL, SDW, and CO in the phase diagram. 
\revision{}{The FF being a gapless phase makes it challenging to investigate in a finite size study, and thus we hereby focus mainly on the qualitative signature of FF.}

The signature of this phase lies in its electron momentum distribution, $n_k = \frac{1}{N}\sum_{ij} e^{-i\vec{k} \cdot \vec{r}_{ij}} \langle c^\dag_i c_j\rangle$. 
The representative data for FF is obtained at $V_1/U\approx 0.0175$ as shown in Fig.~\ref{fig:ff}.
At $U/t=7.0$, which is in the FF phase, we observe a rapid drop in occupation at $q_x/\pi=0.75$, resembling a finite residue similar to a Fermi liquid  (for larger $U/t\gtrsim 12$ in FF, the nature of the liquid phase becomes more complex and difficult to be characterized, which we leave for  future studies).  In contrast, other phases show continuous changes in $n_k$ through the Brillouin zone. 


In the large $U/t$ region, increasing further-range interactions parameterized by $V_1/U$ melts the SDW into FF and then drives the formation of CO. 
Fig.\ref{fig:large_u}(a) shows the spin structure factor $\sq$ along a cut in the momentum space. 
We observe that the amplitude of the $\sq$ peaks decreases as $V_1/U$ increases, and ultimately the peaks disappear at $V_1/U \approx 0.2$.
Increasing the further-range interaction strength across $V_1/U \approx 0.2$, we see the FF until the CO emerges at $V_1/U\approx 0.5$.
We identify the existence of CO from the charge density with a shift to remove amplitude at zero wave vector, $N_k = \frac{1}{\sqrt{N}}\sum_i (n_i-1)e^{-i\vec{k} \cdot \vec{r}}$. 
As shown in Fig.~\ref{fig:large_u}(b), peaks in $N_k$ appear at two corners of the Brillouin zone, corresponding to CO wavevector ${\mathbf Q}_\text{CO}=\left( 0, \pi/a\right)$, where $a$ is the lattice constant. 
Alternatively, in Fig.~\ref{fig:large_u}(c), we show the charge density in real space. 
Given the circular Fermi surface(see Supplemental Materials~\cite{sm} Appendix. E) and wave vector far from $2k_F$, the CO is not driven by nesting or the quasi 1D geometry. Rather, as the strong coupling analysis in Appendix. E suggests the observed CO is driven by strong coupling physics, as in the generalized Wigner crystal phenomena so far only observed at fractional filling\cite{Regan2020Nature, Li2021Nature, Xu2020Nature}.

{\it Conclusion and outlook --} In summary, we investigated the triangular lattice Hubbard model with an extended range of interactions, as motivated by recent experimental developments in hetero TMD Moir\'e systems. Specifically, we explored the two-dimensional phase diagram controlled by the electron hopping $t$ and the further-range interactions $V_1$. 
In the small $V_1/U$ limit, we reproduce results from \cite{Shirakawa2017Phys.Rev.B, Szasz2020Phys.Rev.X, Chen2021} showing transitions from a metallic state to a chiral spin liquid and then to a spin density wave phase as $U/t$ increases.
Upon increasing long-range interactions $V_1, V_2$ and $V_3$, we find that 
the chiral spin liquid is strengthened with small long-range interaction before it gives way to the metallic state. On the other hand, the spin density wave is continuously weakened by increasing long-range interactions. Further increasing further-range interactions, the CO emerges with the wavevector that is not dictated by the quasi-1d geometry of the simulation but rather has periodicity in the direction perpendicular to the cylinder long direction (the $x$ direction as shown in Fig.~\ref{fig:large_u}).
The enhancement of chiral correlation with small but finite further-range interaction we observed presents a tantalizing potential. While we cannot rule out the finite size effect, one possible mechanism for such phenomena would be that further-range interaction-driven direct exchange can suppress SDW ordering and further frustration. The observation warrants further computational and experimental exploration.
The MIT between the interaction-driven CO state and the FF has the potential of supporting a superconducting state that preempts the MIT, similar to the CSL phase preempting the direct phase transition between FF and SDW. Given that the CO state has a localized doubly occupied site, introducing phase coherence and liberating the pairs to move could result in a pair density wave state\cite{Peng2021} in an experimentally realizable setting. Our preliminary results support such a possibility, which is an interesting direction for future investigation\cite{zhou2022inprep}.
In general, the rich phase diagram we uncovered upon tuning the further-range interactions can guide experimental exploration of the new solid-state quantum simulator platform of hetero TMD Moir\'e systems. 

\noindent {\bf Acknowledgements:} 
The authors thank Kin Fai Mak and Jie Shan for helpful discussions. 
Part of the DMRG calculation uses the \mbox{ITensor} package\cite{itensor}. DMRG simulations at Cornell were carried out on the Red Cloud at the Cornell University Center for Advanced Computing, with the support of the DOE under award DE-SC0018946.
EAK and YZ acknowledge support by the National Science Foundation  through award \#OAC-1934714 (Institutes for Data-Intensive Research in Science and Engineering – Frameworks) and through 
the NSF MRSEC program (DMR-1719875) 
for the initial design of the studies.
YZ and E-AK acknowledge support from the Cornell College of Arts and Sciences through the New Frontier Grant. This research is funded in part by the Gordon and Betty Moore Foundation through Grant GBMF10436 to E-AK to support the work of YZ.
DNS acknowledges the support by the  U.S. Department of Energy, Office of Basic Energy Sciences under Grant No. DE-FG02-06ER46305.


\bibliography{bibliography}
\clearpage
\input{sup}

\end{document}

%% file: sup.tex
\onecolumngrid
\begin{center}\
\textbf{\large Supplemental Materials: Quantum Phases of Transition Metal Dichalcogenide Moir{\'e} Systems}
\end{center}\
\twocolumngrid
\appendix
\section{Derivation of $V_2/V_1$ and $V_3/V_1$}
\label{Appendix:gate_geometry}
\begin{figure}[htbp]
  \centering
  \includegraphics[width=0.8\columnwidth]{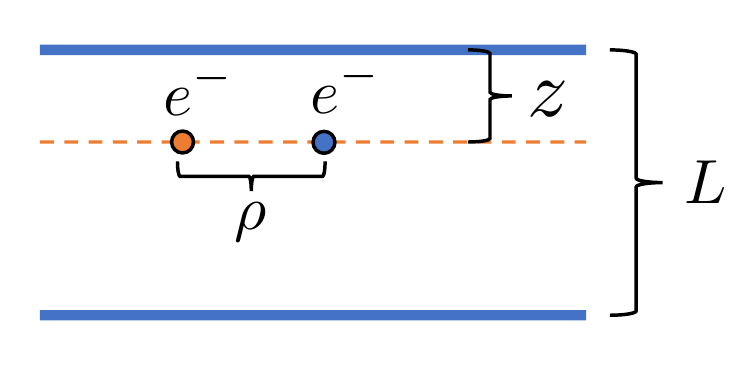}
  \caption{
    Parallel capacitor model for TMD placed between double gates. 
    The figure is drawn from a side view, with lines representing planes extending perpendicular to the page.
    The blue lines represent the top and bottom gates, and the dashed orange line represents the TMD layer.
    Two electrons in the TMD, represented by the circles, are separated by a distance $\rho$. 
    The distance between TMD and the top gate is $z$, and the gate separation is $L$. 
  }
  \label{fig:parallel_capacitor}
\end{figure}

This appendix shows the derivation of the $V_2/V_1$ and $V_3/V_1$ ratios used in our calculation. We model the double gating setup used in TMD experiments as a parallel capacitor, as shown in Fig.~\ref{fig:parallel_capacitor}. 
The long-range inter-site interaction strength is viewed as the potential energy between two electrons separated by distance $\rho$ lying in the TMD layer, which is a plane parallel to the gates. 
The distance between the plane to the top gate is $z$, and the gate separation is $L$.
Solving the boundary value problem, we get the potential energy between the two electrons to be 
\begin{equation}
  V(\rho, z) = \frac{e^2}{\pi \epsilon_0 L} \sum_{n=1}^{\infty} \sin \left( \frac{n \pi z}{L}\right) \sin \left( \frac{n \pi z_0}{L}\right) \mathbf{K_0}\! \left( \frac{n\pi \rho}{L}\right),
\end{equation}
where $\mathbf{K_0}$ is the modified Bessel function. 
In the main text, we present results of a specific gate-sample geometry where the sample is placed $5$nm away from the top gate ($D_t=5$nm) and $30$nm away from bottom gate ($D_b=30$nm), corresponding to $z=5$ nm and $L=35$ nm. A typical Moir{\'e} period is $a_M=5$ nm. 
On a triangular lattice, nearest-neighbor interactions have $\rho_1=a_M$, second nearest neighbors $\rho_2=\sqrt{3}a_M$ and third nearest neighbors $\rho_3=2a_M$. 
Combining all these data, the model gives $V_2/V_1\approx 0.357$ and $V_3/V_1\approx 0.260$.

\section{Effects of variations in device geometry on COS}
\begin{figure}[htbp]
  \centering
  \includegraphics[width=0.8\columnwidth]{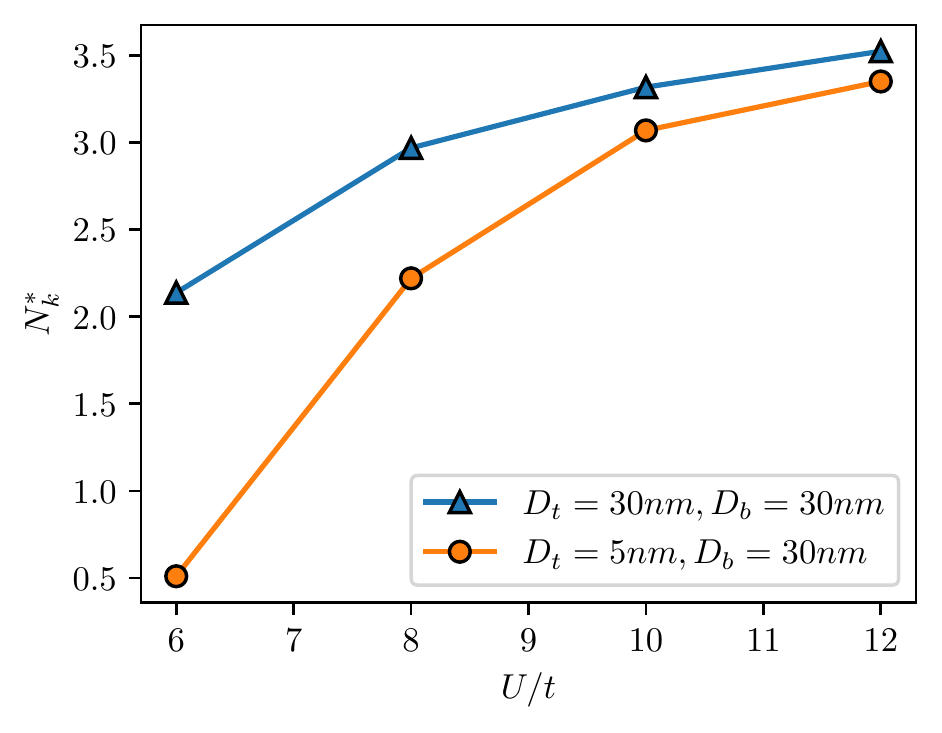}
  \caption{
    COS intensity $N_k^*$ as a function of the device geometry and repulsive interaction strength. All calculations done at $V_1/U \approx 0.56$.
  }
  \label{fig:two_gate_geometries}
\end{figure}

As captured by this model, one can change the gate sample separation to tune the screening of Coulomb interaction and effectively adjust $V_2/V_1$ and $V_3/V_1$. In small $V_1/U$ limit, all further-range interactions are weak and the standard Hubbard model behavior should be retrieved as $V_1/U \rightarrow 0$. Therefore, we focus on the effect of $V_2/V_1$ and $V_3/V_1$ in the large $V_1/U$ region.
In Fig.\ref{fig:two_gate_geometries}, we compare two sets of $V_2/V_1$ and $V_3/V_1$ derived from two experimentally practical gate geometries (Table.~\ref{table}). To isolate the effect of $V_2$ and $V_3$, we fix $V_1/U \approx 0.56$ and vary $V_2/V_1$ and $V_3/V_1$. As shown in  Fig.\ref{fig:two_gate_geometries}, the $D_t=5$nm, $D_b=30$nm case, which has weaker screening effect and larger $V_2/V_1$ and $V_3/V_1$ than the geometry considered in main text, promotes the COS phase. Therefore, we conclude that in a considerable range of $V_2/V_1$ and $V_3/V_1$, all phases in the phase diagram (main text Fig.1(b)) can be observed. The exact position of phase boundary might be distorted by $V_2/V_1$ and $V_3/V_1$.

\begin{table}[h]
\begin{tabular}{|c|c|c|}
\hline
Gate geometry         & $V_2/V_1$    & $V_3/V_1$    \\ \hline
$D_t=5$nm, $D_b=30$nm & $0.35676654$ & $0.26046785$ \\ \hline
$D_t=30$nm, $D_b=30$nm & $0.52357966$ & $0.43674471$ \\ \hline
\end{tabular}
\caption{
$V_2/V_1$ and $V_3/V_1$ values corresponding to different gate geometries. }
\label{table}
\end{table}

\section{Convergence of DMRG calculations }
\label{Appendix:DMRG_convergence}
\begin{figure}[htbp]
  \centering
  \includegraphics[width=0.8\columnwidth]{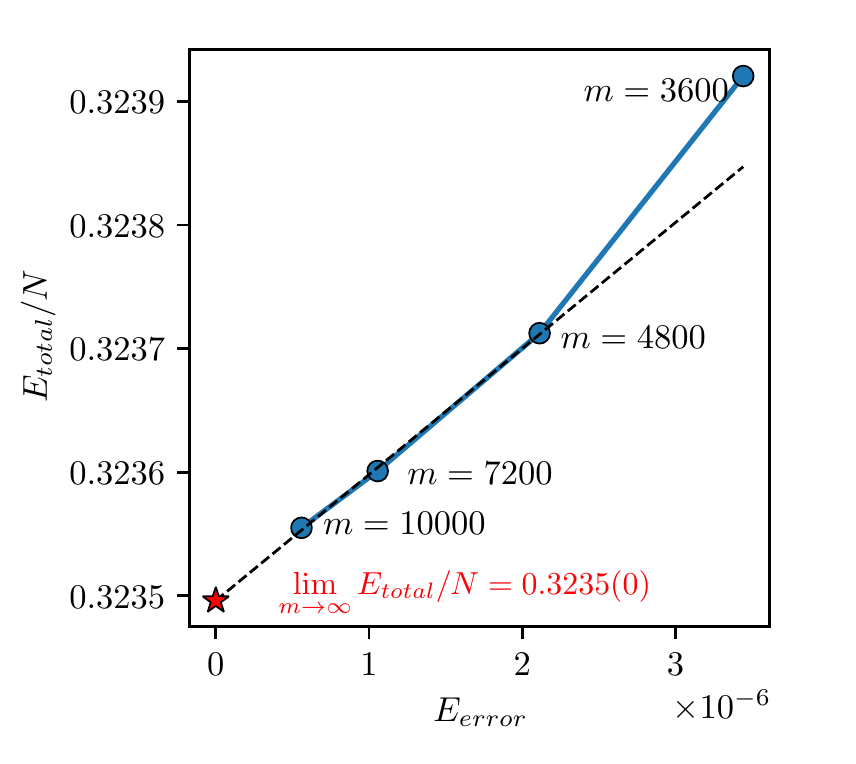}
  \caption{
    Convergence of energy. Energy per site $E_{total}/N$ versus truncation error $E_{error}$. Corresponding bond dimensions $m$ for SU(2) DMRG are labeled.The dashed line extrapolates $E_{total}/N$ as a function of $E_{error}$. In the limit $E_{error}\rightarrow 0, E_{total}/N\rightarrow0.3235(0)$ based on the extrapolation.
  }
  \label{fig:energy_convergence}
\end{figure}

\begin{figure}[htbp]
  \centering
  \includegraphics[width=0.98\columnwidth]{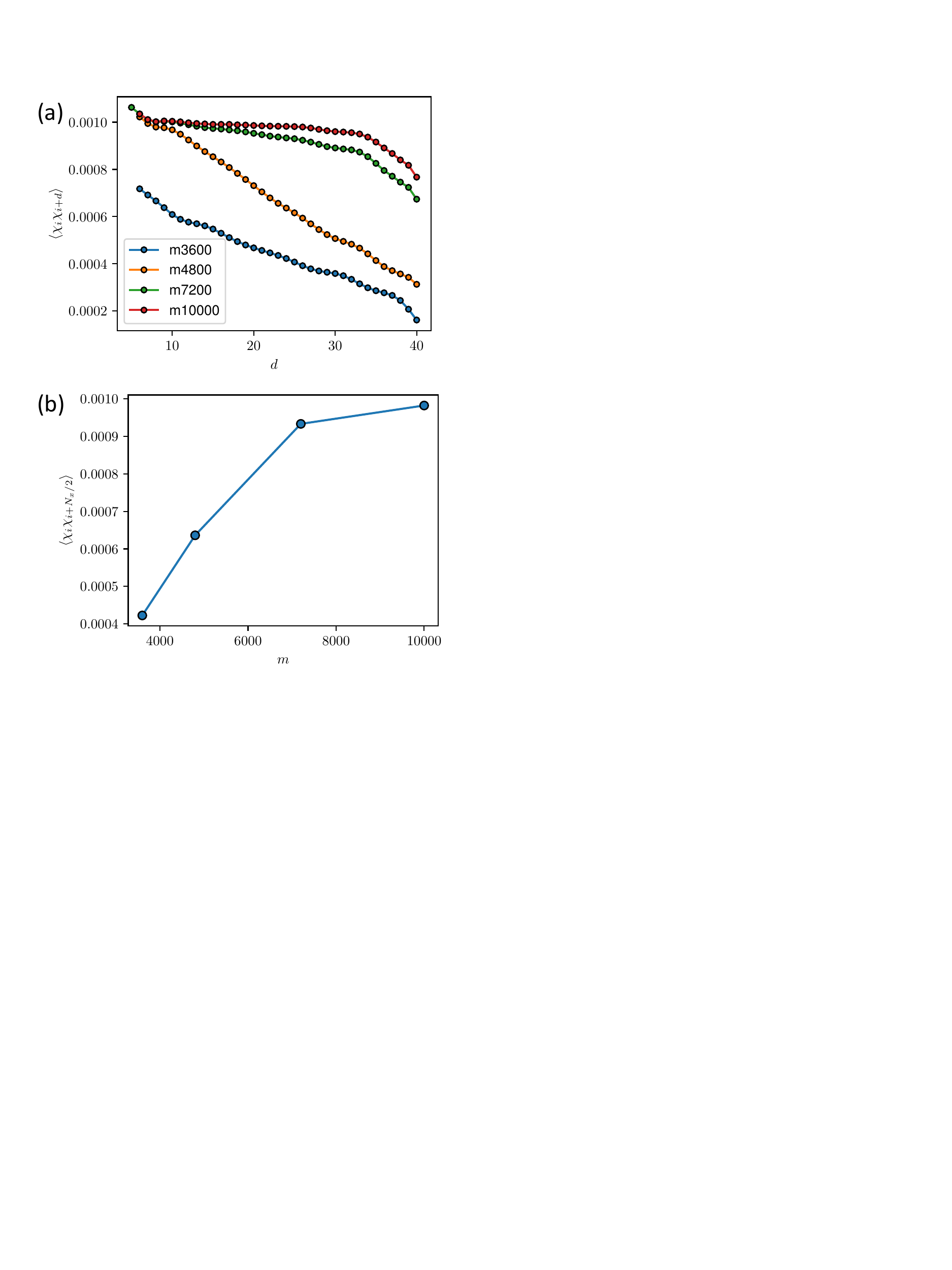}
  \caption{
    Convergence of chiral correlation function. (a) Chiral correlation $\langle \chi_{i} \chi_{i+d}\rangle$ as a function of site separation $d$ at different bond dimensions $m$. (b) Chiral correlation at half cylinder site separation $\langle \chi_{i} \chi_{i+N_x/2}\rangle$ versus bond dimension $m$ in SU2 DMRG. Calculation performed on YC$4$ cylinders with $N_x = 48$.
  }
  \label{fig:csl-convergence}
\end{figure}

In this appendix, we provide evidence of the convergence of the DMRG calculations. 
As shown in Fig.~\ref{fig:energy_convergence}, the relative error in energy decreases as the number of states (reflected by the bond dimension $m$) increases. We keep $m=10000$ SU(2) multiplets in DMRG (which is about $30,000$ U(1) states), the calculation is well converged with the truncation error on the order of $10^{-6}$. 
Fg.~\ref{fig:csl-convergence} shows further details on the convergence of chiral correlation strength, which is the crucial indicator of the existence of the CSL phase. In Fig~\ref{fig:csl-convergence}(a), we see that the system builds up long-range chiral order as the number of kept states increase. As shown in Fig.~\ref{fig:csl-convergence}(b), the chiral correlation strength at half-cylinder length is stabilized at a large bond dimension.  Based on the enhancement of chiral correlation with increasing bond dimension, we conclude that the state will be a CSL state presenting long-range chiral order in the limit $m \rightarrow \infty$. 

\section{Correlation functions in the real space}
\label{Appendix:real_space_correlation}
In this appendix, we show correlation functions viewed in the real space. As a supplement to the structure factors discussed in the main text, the real space single-particle Green's function and the spin correlation functions provide further insight into the phase transition from FF to CSL and CSL to SDW correspondingly.

\begin{figure*}[t!]
  \centering
  \includegraphics[width=1.5\columnwidth]{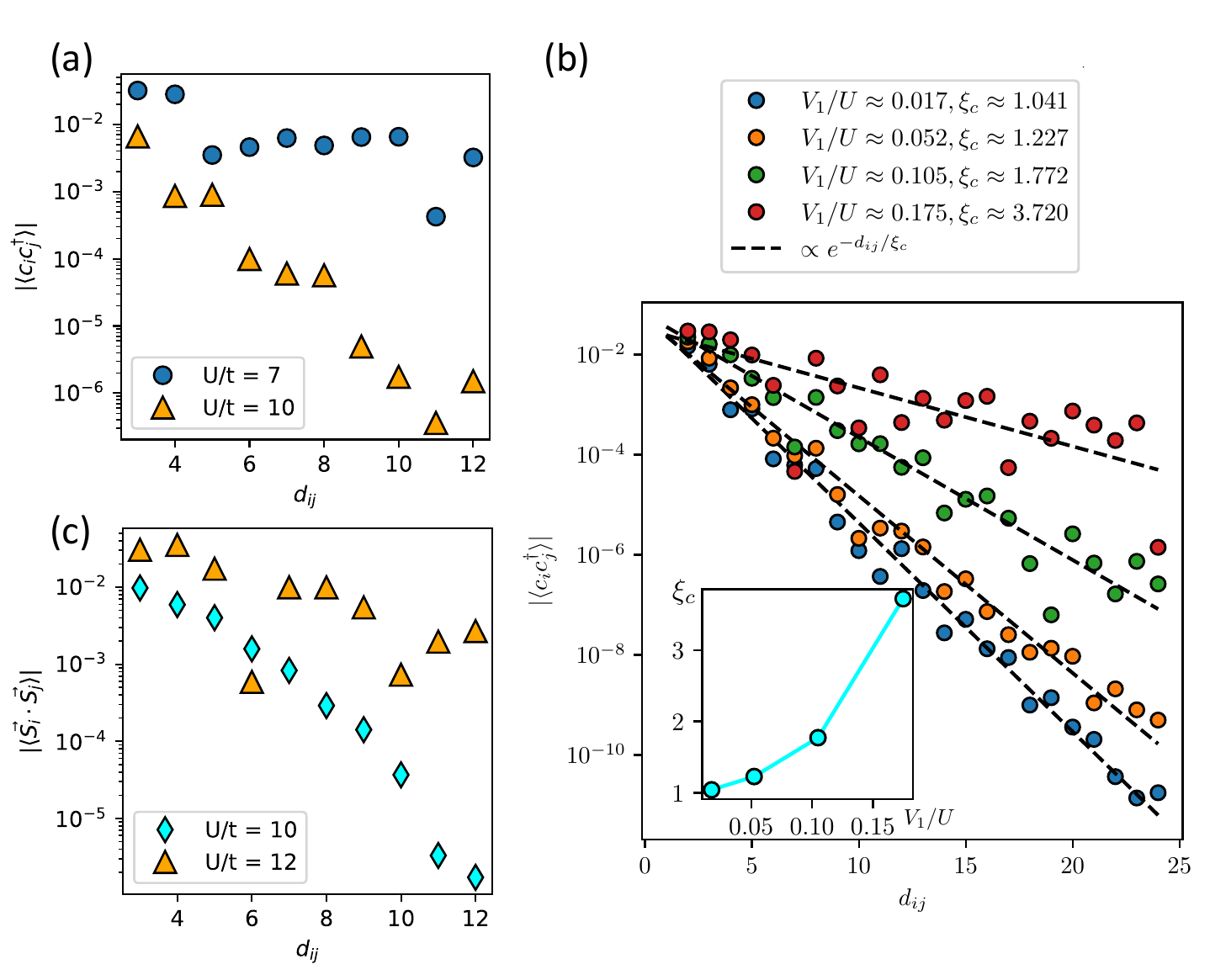}
  \caption{ Correlation function in real space. (a): Single particle Greens function $|\langle c_i c_j^\dag\rangle|$ versus site separation $d_{ij}$ for $U/t=7$ (FF) and $U/t=10$ (CSL) at  $V_1/U \approx 0.017$. 
  (b):Main panel: Single particle Green's function $|\langle c_i c_j^\dag \rangle|$ versus distance $d_{ij}$ for various $V_1/U$ at $U/t=10$. The markers are calculated correlation strength, while the dashed lines are data fitted to scaling function of form $\propto e^{-d_{ij}/\xi_c}$.The extrapolated correlation length $\xi_c$ values are listed in the legend. Inset: Extrapolated correlation length $\xi_c$ as a function of the further range interaction strength parameterized by $V_1/U$.
  (c): Spin correlation function $\langle S_i \cdot S_j\rangle$ versus site separation $d_{ij}$ for $U/t=10$ (CSL) and $U/t=12$ (SDW) at $V_1/U \approx 0.017$. 
  }
  \label{fig:real-space-cor}
\end{figure*}

For the transition from FF to CSL we look at the single-particle Green's function $|\langle c_i c_j^\dag\rangle|$. The interplay of $U/t$ and $V_1/U$ leads to a dome-shaped phase boundary between the FF and CSL phases. We first focus on a horizontal cut in the phase diagram at fixed $V_1/U\approx0.017$. As shown in in Fig.~\ref{fig:real-space-cor}(a), at $U/t=7$, corresponding to the FF phase, $|\langle c_i c_j^\dag\rangle|$ persists even for large site separation. For $U/t=10$, corresponding to the CSL phase, the correlation strength decays exponentially with respect to the site separation $d_{ij}$. The observation of exponential decay or not in $|\langle c_i c_j^\dag\rangle|$ draws the difference between the FF and CSL. We further probe the correlation length $\xi_c$ dependence on $V_1/U$ along a vertical cut in the phase diagram at $U/t=10$ for a range of $V_1/U$ that crosses the phase boundary between FF and CSL in ~Fig.\ref{fig:real-space-cor}(b). We observe a short correlation length in the CSL phase, corresponding to a small $V_1/U$ limit. In contrast, for $V_1/U = 0.175$, which is deep into the FF phase, we observe a much longer correlation length. As the further range interaction strengthens, the correlation length shows a significant increase, which is consistent with our observation of CSL melted into FF by further range interactions.
To study the transition from CSL to SDW, we look at the spin correlation function in the real space, as presented in Fig.~\ref{fig:real-space-cor}(c). The spin correlation decays exponentially at $U/t=10$, corresponding to the CSL phase. In comparison, the SDW phase shows a long-range spin correlation.

\section{Interaction-driven charge order}
We investigate the nature of the charge-ordered phase observed at large $U/t$ and large $V_1/U$ (see Fig.1(b) in the main text) by considering the kinetics and energetics; we show that the charge-ordered state is interaction-driven.

\begin{figure}[h!]
  \centering
  \includegraphics[width=0.7\columnwidth]{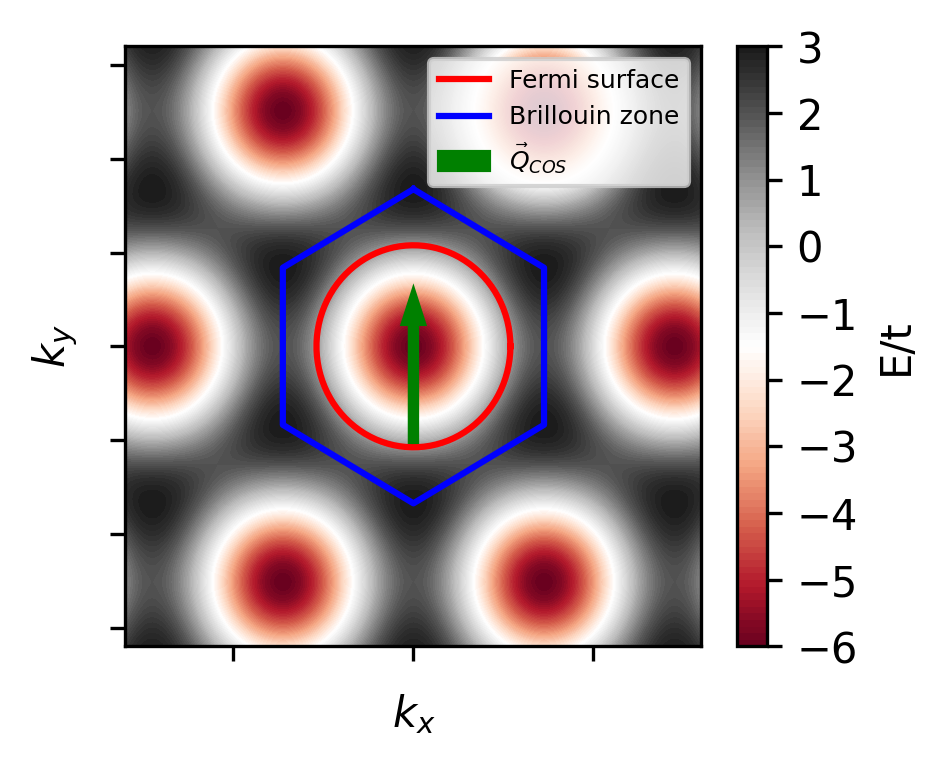}
  \caption{ Band structure from a tight-binding model on the triangular lattice. The 1st Brillouin zone (blue) and the Fermi surface at half-filling (red) are marked. The green arrow is the charge order wavevector.
  }
  \label{fig:nesting}
\end{figure}

\begin{figure}[]
  \centering
  \includegraphics[width=0.9\columnwidth]{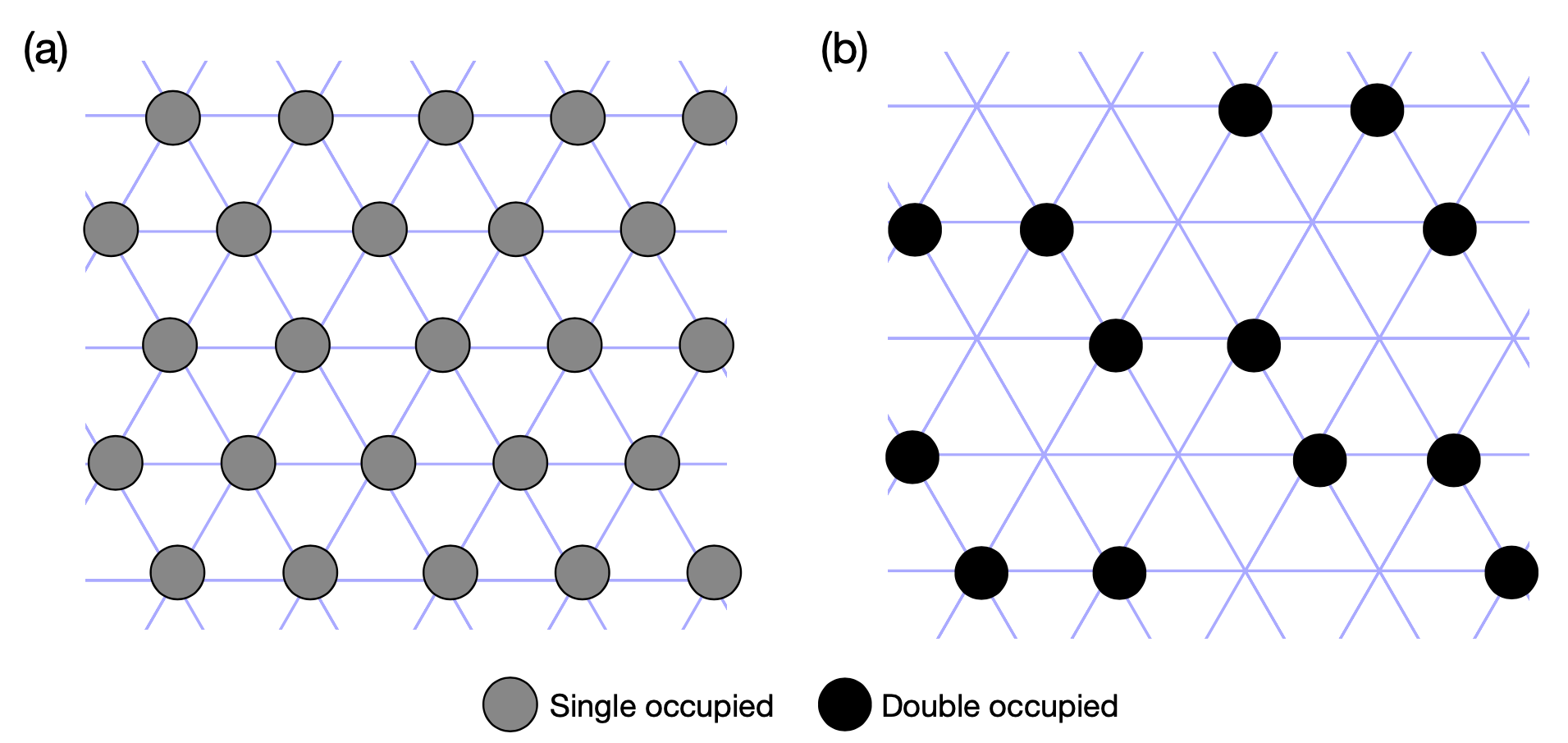}
  \caption{
    Schematics of two charge orderings. Panel (a) resembles the Mott insulating state with one electron occupying each site. Panel (b) resembles the charge order observed from the DMRG simulation. Double-occupied sites are marked with dark dots; all other sites are unoccupied.
  }
  \label{fig:charge-order}
\end{figure}

We first show concrete evidence that this charge-ordered state is not a conventional nesting-driven charge density wave by considering a tight-binding model on the triangular lattice. As shown in Fig .\ref{fig:nesting}, the wavevector observed from the DMRG simulation does not connect different portions of the Fermi surface, causing no nesting. With the possibility of Fermi surface instability ruled out, we focus on the Coulomb interaction part of the Hamiltonian 
\begin{equation}
    H_{\text{int}} =   U\sum_{i} n_{i, \uparrow} n_{i, \downarrow} 
     + V_1 \sum_{\left<ij\right>} n_i n_j + V_2 \sum_{\langle\!\langle ij \rangle\!\rangle } n_i n_j + V_3 \sum_{\langle\!\langle\!\langle ij\rangle\!\rangle\!\rangle} n_i n_j,
\end{equation}
Specifically, competition between two charge orderings at half-filling, sketched in Fig.\ref{fig:charge-order}, is studied. The energy of configuration Fig.~\ref{fig:charge-order}(a) is 
\begin{equation}
    E_{\text{Mott}} = 3N \cdot(V_1+V_2+V_3).
\end{equation}
The energy of Fig.~\ref{fig:charge-order}(b) is  
\begin{equation}
    E_{\text{stripe}} = N \cdot(\frac{U}{2}+2V_1+4V_2+2V_3),
\end{equation}
where $N$ is the number of particles.
The energy difference between these two configurations is
\begin{equation}
    \Delta_E \equiv E_{\text{stripe}} - E_{\text{Mott}}
    = N\left( U/2 -V_1 + V_2 -V_3\right). 
\end{equation}
For the gate geometry considered in the main text, we have $\Delta_E= N \cdot \left( U/2 - 0.9037 V_1 \right)$ from which one expects that Fig.~\ref{fig:charge-order}(b) will be favorable in energy when $V_1/U \gtrsim 0.55328$ in thermodynamic limit. This estimate roughly matches with the DMRG prediction for the CO in the phase diagram (see main text Fig.1(b)) and thus we conclude that this charge-ordered state is driven by the Coulomb interactions.

